\documentclass[aps,column,print]{revtex4}
\usepackage{amsfonts}
\usepackage{amsmath}
\usepackage{amssymb}
\usepackage{graphicx}
\usepackage{color}

\begin{document}

\title{Two-dimensional quantum droplets in binary quadrupolar condensates}
\author{Aowei Yang$^{1}$, Jiahao Zhou$^{1}$, Xiaoqing Liang$^{1}$, Guilong Li%
$^{1}$}
\author{Bin Liu$^{1,2}$}
\email{binliu@fosu.edu.cn}
\author{Huan-Bo Luo$^{1,2,3}$}
\email{huanboluo@fosu.edu.cn}
\author{Boris A Malomed$^{4,5}$}
\author{Yongyao Li$^{1,2}$}
\affiliation{$^{1}$School of Physics and Optoelectronic Engineering, Foshan University,
Foshan 528000, China}
\affiliation{$^{2}$Guangdong-Hong Kong-Macao Joint Laboratory for Intelligent Micro-Nano
Optoelectronic Technology, Foshan University, Foshan 528000, China}
\affiliation{$^3$Department of Physics, South China University of Technology, Guangzhou
510640, China}
\affiliation{$^{4}$Department of Physical Electronics, School of Electrical Engineering,
Faculty of Engineering, and Center for Light-Matter Interaction, Tel Aviv
University, Tel Aviv 69978, Israel}
\affiliation{$^{5}$Instituto de Alta Investigaci\'{o}n, Universidad de Tarapac\'{a},
Casilla 7D, Arica, Chile}

\begin{abstract}
We study the stability and characteristics of two-dimensional (2D)
quasi-isotropic quantum droplets (QDs) of fundamental and vortex types,
formed by binary Bose-Einstein condensate with magnetic
quadrupole-quadrupole interactions (MQQIs). The magnetic quadrupoles are
built as pairs of dipoles and antidipoles polarized along the $x$-axis. The
MQQIs are induced by applying an external magnetic field that varies along
the $x$-axis. The system is modeled by the Gross-Pitaevskii equations
including the MQQIs and Lee-Huang-Yang correction to the mean-field
approximation. Stable 2D fundamental QDs and quasi-isotropic vortex QDs with
topological charges $S\leq 4$ are produced by means of the
imaginary-time-integration method for configurations with the quadrupoles
polarized parallel to the system's two-dimensional plane. Effects of the
norm and MQQI strength on the QDs are studied in detail. Some results,
including an accurate prediction of the effective area, chemical potential,
and peak density of QDs, are obtained in an analytical form by means of the
Thomas-Fermi approximation. Collisions between moving QDs are studied by
means of systematic simulations.
\end{abstract}

\maketitle

\section{Introduction}

In 2015, Petrov had proposed a scheme for suppression of the collapse in the
binary (two-species) Bose-Einstein condensate (BEC) in the three-dimensional
(3D) space with $g_{12}^{2}>g_{11}g_{22}$, where $g_{12}$ is the strength of
the inter-species attraction, while $g_{11}$ and $g_{22}$ are strengths of
the intra-species self-repulsion \cite{petrov2015QDs}. The stabilization of
the system against the collapse is provided by the Lee-Huang-Yang (LHY)
effect, i.e., corrections to the mean-field BEC dynamics induced by quantum
fluctuations \cite{LHY}. The analysis had predicted the arrest of the
collapse and formation of stable quantum droplets (QDs) filled by the
ultra-diluted quantum fluid. Following the prediction, QDs have been created
in a mixture of two hyperfine atomic states of $^{39}$K, with the quasi-2D
\cite{QD_K,QD_Tarruell_2} and fully 3D \cite{PRL120_235301} shapes, as well
as in a heteronuclear mixture of $^{23}$Na and $^{87}$Rb \cite{QD_KNA_WDJ}
atoms. The use of the LHY effect has also made it possible to create QDs in
single-component dipolar bosonic gases of dysprosium \cite{QD_dy_nature} and
erbium \cite{QD_er_prx} atoms. The further analysis has demonstrates that
the LHY corrections take different forms in 1D, 2D, and 3D configurations
\cite{petrov2016}.

Studies of QDs have drawn much interest in very different contexts \cite%
{PRA103_033312,PRL120_160402,PRA98_033612,PRA_HH,PRR4_013168,PRA102_023318,
dong2022bistable,PRA103_013312,PRE102_062217,PRA101_051601(R),FOP_LZH,CXL_PRA2018, PRResearch2_043074,PRL126_244101,CSF_WHC,FOP_GMY,FOP_Boris,Nonlinear Dyn. Zhou 2022, CSF Zhou 2021,NJP1,NJP5,NJP6,NJP7,NJP8,cpb5,cpb_yw,cpb_lxj,PRl_122_090401,CSF_ZFY,RPP_dipolar,PRS,JPBmagnetic, Fop_zhangyb,zhangYC2018long,zhangYC2023quantum,zhangYC2021self,Yin2021,chen2021one, ma2021borromean,cui2021droplet,xu2022three,zhou2019dynamics,wang2020thermal, boudjemaa2023quantum,pshenichnyuk2017static,rakshit2019self,zin2022self,Fischer2006}%
. Various phenomena, such as the spin-orbit coupling \cite{LYY_SOC_QD},
supersolidity \cite{prx_Supersolid,zhangYC2021phases}, and metastability
\cite{metastability} can be emulated in BECs under the action of the LHY
corrections. Further, stable semi-discrete vortex QDs were predicted in
arrays of coupled 1D cigar-shaped traps \cite{PRL_LYY2019}, and the symmetry
breaking of QDs has been studied in a dual-core trap \cite{LB_PRA}.

In particular, an interesting possibility is to create stable solitary
vortices, i.e., QDs with embedded vorticity (traditional quantized vortices
in BEC are eddy states supported by a flat background, i.e., they are 2D
dark solitons \cite{Fetter}, while solitary vortices resemble bright
matter-wave solitons). Such modes are often subject to the azimuthal
modulational instability that develops faster than the collapse, splitting
the vortices into fragments. Therefore, stability conditions for vortex QDs
become increasingly stricter with the increase of their winding number
(alias the topological charge). The possibility for the stabilization of the
vortex modes has been revealed by studies of BEC systems under the action of
spatially periodic and quasiperiodic lattice potentials \cite{BEC_OL, cpb3,
cpb1, cpb_wlh, cpl_dcq, cpl_xsl, RFZ2, WJG, ZXFPRA95, Brtka, HV_WB, HV_pla,
PRA_HCQ2017}. Recently, stable QDs in 2D square lattices \cite{FOP_ZYY} and
ring-shaped QDs with explicit and hidden vorticity in a radially periodic
potential \cite{NJP_LB2022,PRE_LB} have been predicted (\textquotedblleft
hidden vorticity" means as a vortex-antivortex bound state in a
two-component system with zero total angular momentum \cite{Brtka}). In the
free space, stable 2D matter-wave solitons were predicted in
microwave-coupled binary condensates \cite{QJL_PRA}, and solutions for
stable semi-vortices, i.e., 2D \cite{Ben Li} and 3D \cite{Han Pu}
spin-orbit-coupled two-component matter-wave solitons carrying vorticity in
one component, are known too.

In dipolar BECs, QDs have been observed in the form of 3D self-bound states.
In the 2D geometry, the coherence of QDs was explored in Ref. \cite%
{2Ddipolar_njp}. In Ref. \cite{pragroundstate}, the ground-state properties
and intrinsic excitations of QDs in dipolar BECs were studied in detail. The
previous studies demonstrate that QDs in dipolar BECs feature strong
anisotropy of their density profile in the free space, but they do not carry
vorticity. In Ref. \cite{dip_PRA-unstable}, isotropic vortex QDs with the
dipoles polarized parallel to the vortical pivot were constructed and found
to be completely unstable. Thus, creating stable vortex QDs in BECs with
dipole-dipole interactions (DDI) remains a fundamental problem. Recently,
solutions for 2D fundamental QDs in dipolar BEC with the dipoles oriented
perpendicular to the system's plane have been produced \cite{photonics_YAW}.
Furthermore, Li\textit{\ et al}. have proved that dipolar BEC offers a
unique possibility to predict stable 2D anisotropic vortex QDs with the
vortex axis oriented perpendicular to the polarization of dipoles \cite%
{FOP_LGL}. Hence, modifying the attractive nonlinearity by introducing DDI
is a promising direction for the stabilization of vortex QDs in BEC. The
objective of the present work is to demonstrate that the stabilization is
also possible with the help of long-range quadrupole-quadrupole interactions
(QQI)\ between particles in BEC. Previously, the formation of solitons in
QQI-coupled BEC was addressed in models that did not include the LHY terms
\cite{QQI_LYY_PRA,FOP_QQI_LYY,ZRX_QQI, mishra2020self,ghosh2022droplet},
which made it more difficult to stabilize the self-trapped states in the BEC.

Electric quadrupoles may be built as tightly bound pairs of dipole and
antidipoles directed perpendicular to the system's ($x$, $y$) plane, i.e.,
along the $z$-axis. By applying an external gradient electric field along
the $z$-axis the electric QQI (alias EQQI) can be induced. The potential of
the interaction between two electric quadrupoles in this configuration is
written in the polar coordinates $\left( r,\theta \right) $ as%
\begin{equation}
U_{QQ}^{(\mathrm{elect})}(r,\theta )=\frac{4}{3}Q^{2}r^{-5}\left( 3-5\cos
^{2}\theta \right) ,  \label{QQI_elec}
\end{equation}%
where $Q$ is the quadrupole moment, $r$ is the distance between the
quadrupoles, and $\theta $ is the angle between the vector connecting the
quadrupoles and the line connecting the dipole and antidipole inside the
quadrupole. Averaging potentials (\ref{QQI_elec}) over the angular range $%
0\leq \theta <2\pi $ yields the respective mean values:
\begin{equation}
\frac{1}{2\pi }\int_{0}^{2\pi }U_{QQ}^{(\mathrm{elect})}\left( r,\theta
\right) d\theta =-2Q^{2}r^{-5}<0.  \label{QQI_elec_integral}
\end{equation}%
According to Eq. (\ref{QQI_elec_integral}), the EQQI represent the
attractive interaction.

Relatively large electric quadrupole moments are featured by some small
molecules \cite{molecules} and by bound states in the form of alkali diatoms
\cite{diatoms}. These particles can be used to create BEC in which EQQI
plays an essential role. In Ref. \cite{FOP_QQI_LYY}, the creation of
quadrupolar matter-wave solitons in the 2D free space was predicted, with a
conclusion that, in the presence of EQQI, the solitons feature a higher mass
and stronger anisotropy than their DDI-maintained counterparts, for the same
environmental parameters. Possibilities of building 2D (quasi-) discrete
matter-wave solitons composed of quadrupole particles trapped in deep
isotropic and anisotropic optical-lattice potentials was demonstrated in
Refs. \cite{QQI_LYY_PRA} and \cite{ZRX_QQI}, respectively.\ In Ref. \cite%
{CGH_QQI_CNCNS}, soliton solutions of the mixed-mode and semi-vortex types
were constructed in binary quadrupolar BECs including spin-orbit coupling.

Magnetic quadrupoles can be built as tightly bound dipole-antidipole pairs
directed along a particular axis ($x$). They can be polarized parallel to
each other by a gradient magnetic field also directed along the $x$-axis.
The potential of the magnetic quadrupole-quadrupole interactions (MQQI) is

\begin{equation}
U_{QQ}^{(\mathrm{magn})}(r,\theta )=\frac{4}{3}Q^{2}r^{-5}\left( 3-30\cos
^{2}\theta +35\cos ^{4}\theta \right) ,  \label{QQI_magn}
\end{equation}%
cf. Eq. (\ref{QQI_elec}). Unlike EQQI, which is attractive on the average,
for MQQI the mean value of potential (\ref{QQI_magn})\ corresponds to
repulsion:%
\begin{equation}
\frac{1}{2\pi }\int_{0}^{2\pi }U_{QQ}^{(\mathrm{magn})}\left( r,\theta
\right) d\theta =\frac{3}{2}Q^{2}r^{-5}>0,  \label{QQI_magn_integral}
\end{equation}%
cf. Eq. (\ref{QQI_elec_integral}).

In the usual mean-field approximation, which is represented by the
Gross-Pitaevskii (GP) equation for the single-particle wave function $\psi $
\cite{Pit}, BEC cannot maintain localized states with the help of MQQI,
unless an external trapping potential is used Ref. \cite{cpb_yisu_QQI}.
However, the above-mentioned proposal to create QDs in BEC\ by means of the
LHY correction to the mean-field approximation \cite%
{petrov2015QDs,petrov2016} suggests a possibility to predict self-trapped
objects supported by the QQI. Note that the LHY terms take different forms
for different spatial dimensions. In 3D, this term in the corrected GP
equation is proportional to ${|}\psi |^{3}\psi $, with the sign representing
self-repulsion \cite{petrov2015QDs}. In 2D, the combined mean-field-LHY term
is
\begin{equation}
\mathrm{2D~local~term}~=\mathrm{const}\cdot |\psi |^{2}\ln (|\psi
|^{2}/n_{0})\cdot \psi ,  \label{2D}
\end{equation}%
where $\mathrm{const}$\ is a positive coefficient, and $n_{0}$ is the
reference density \cite{petrov2016} (its typical value is $\sim 10^{14}$ cm$%
^{-3}$). Expression (\ref{2D}) implies the attraction at $|\psi |^{2}<n_{0}$%
, and repulsion at $|\psi |^{2}>n_{0}$. Finally, in 1D, the LHY term is
proportional to $-{|\psi |\psi }$, with the sing corresponding to the
self-attraction, which helps to build solitons and QDs in the effectively
one-dimensional condensate \cite{petrov2016}.

In this work, we produce 2D solutions for self-bound states, including ones
with embedded vorticity, in magnetic quadrupolar BEC, taking the
corresponding LHY correction into account. Following the original proposal\
\cite{petrov2015QDs,petrov2016} which derived the correction for the
two-component BEC, we also address the two-component system. In this
connection, it is relevant to stress that the LHY correction to the GP
equation for the dipolar BEC has the same form (in particular, being $\sim {|%
}\psi |^{3}\psi $ in 3D) as in the case of the BEC with contact-only atomic
interactions \cite{smith2021quantum, bisset2021quantum}. Accordingly, the
same is true for the BEC featuring the long-range MQQI in the combination
with the local nonlinearity.

We here consider the 2D model with coordinates $\left( x,y\right) $,
including a gradient magnetic field oriented along the $x$ direction. This
field is induced by a tapered solenoid, as shown schematically in Fig. \ref%
{Fig1}. In this configuration, the MQQI, defined as per Eq. (\ref{QQI_magn}%
), represents self-repulsion, while the local interaction, combining the
mean-field and LHY terms according to Eq. (\ref{2D}) may account for
attraction. The self-trapping of QDs is achieved through competition between
these terms.

\begin{figure}[h]
\includegraphics[width=0.5\columnwidth]{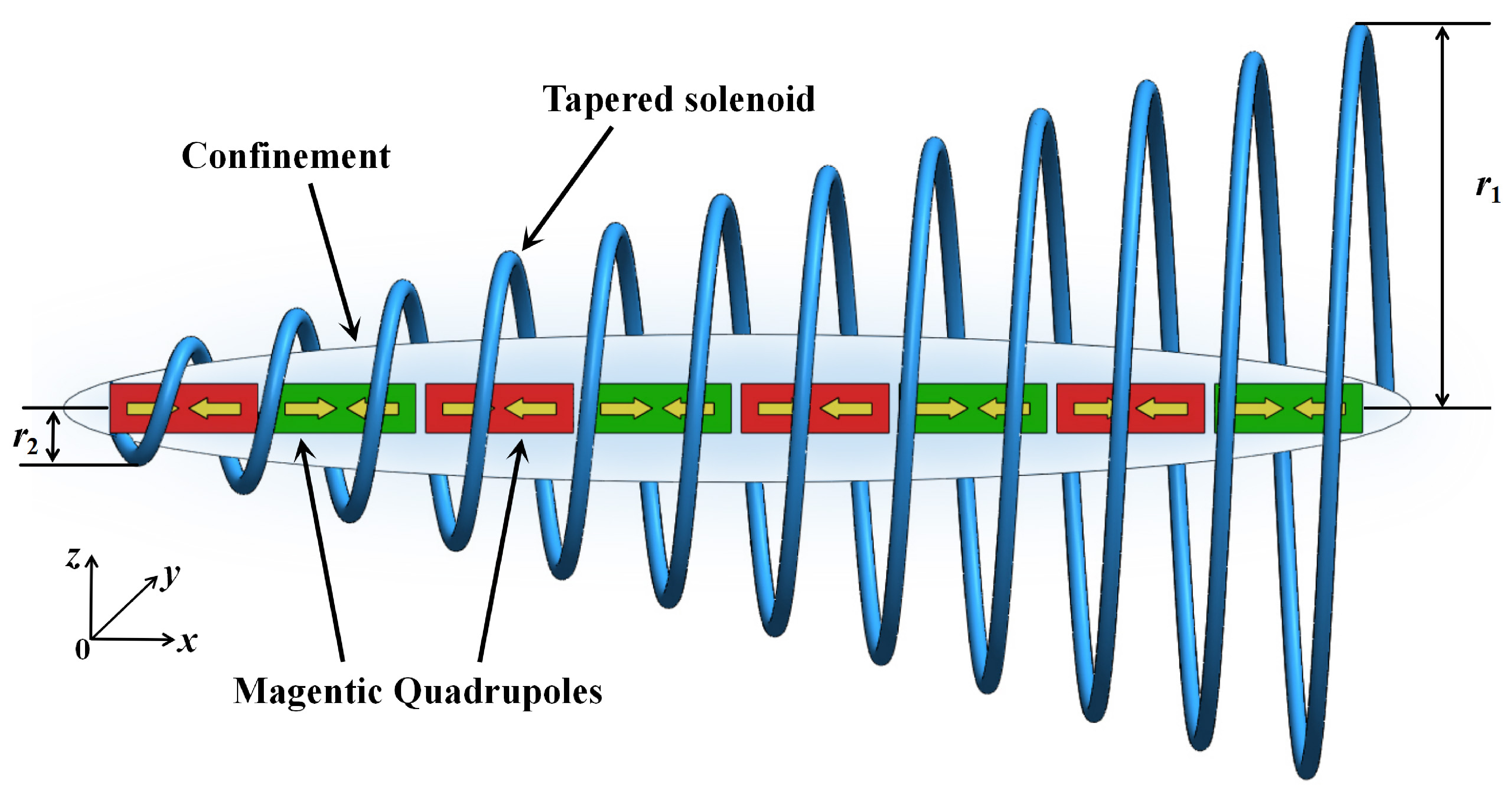}
\caption{The system's scheme. Quadrupoles are built as tightly-bound
dipole-antidipole pairs, polarized along the $x$ axis by means of the
gradient magnetic field (with the local strength being a linear function of $%
x$), cf. Ref. \protect\cite{QQI_LYY_PRA}. This field can be imposed by the
tapered solenoid, as shown in the figure. The expression of the magnetic
field at any point $x_{p}$ belonging to the $x$-axis of the tapered solenoid
is produced by the integration of $dB_{x}\sim \protect\mu %
_{0}nI[r_{2}-(r_{2}-r_{1})x/h]dx/[(r_{2}-(r_{2}-r_{1})x/h)^{2}+(x-x_{p})^{2}]^{3/2}dx
$, where $r_{1}$ and $r_{2}$ are radii of the upper and lower bottom
surfaces of the solenoid, $h$ is the length of the solenoid, $I$ is the
current, and $n$ is the number of turns of the coil. Red and green rectanges
represent two components of the bosonic mixture.}
\label{Fig1}
\end{figure}

Fundamental QDs and quasi-isotropic vortical ones are produced by the
analysis presented below. Effects of the norm and strength of the MQQI are
systematically studied. The rest of the paper is structured as follows. The
model is introduced in Sec. II. Numerical findings and some analytical
estimates are summarized in Sec. III. The dynamics of the QDs are the focus
of Sec. IV. The paper is concluded in Sec. V.

\section{The model}

We aim to consider two-component QDs in the magnetic quadrupolar BEC, which
is made effectively two-dimensional by the application of strong confinement
in the transverse direction, with confinement size $a_{\perp }\sim 1$ $%
\mathrm{\mu }$m. The orientation of individual quadrupoles is fixed as in
Fig. \ref{Fig1}. The system's dynamics are governed by the 2D coupled GP
equations, produced by the reduction of the full 3D GP system. The 2D
equations include the LHY terms, as derived in Ref. \cite{petrov2016}. In
the scaled form, they are written as \cite{FOP_LGL,smith2021quantum,
bisset2021quantum}%
\begin{eqnarray}
i\frac{\partial \psi _{+}}{\partial t} &=&-\frac{1}{2}\nabla ^{2}\psi
_{+}+\left( {g}_{11}{\left\vert \psi _{+}\right\vert }^{2}+{g}_{12}{%
\left\vert \psi _{-}\right\vert }^{2}\right) \psi _{+}+{\gamma }\psi _{+}%
\left[ \left\vert \psi _{+}(\mathbf{r})\right\vert ^{2}+\left\vert \psi _{-}(%
\mathbf{r})\right\vert ^{2}\right]   \notag \\
&&\times \ln \left[ \left\vert \psi _{+}(\mathbf{r})\right\vert
^{2}+\left\vert \psi _{-}(\mathbf{r})\right\vert ^{2}\right] +\kappa \psi
_{+}(\mathbf{r})\iint d\mathbf{r^{\prime }}R\left( \mathbf{r}-\mathbf{%
r^{\prime }}\right) \left[ \left\vert \psi _{+}\left( \mathbf{r^{\prime }}%
\right) \right\vert ^{2}+\left\vert \psi _{-}\left( \mathbf{r^{\prime }}%
\right) \right\vert ^{2}\right] ,  \label{Model1} \\
i\frac{\partial \psi _{-}}{\partial t} &=&-\frac{1}{2}\nabla ^{2}\psi
_{-}+\left( {g}_{22}{\left\vert \psi _{-}\right\vert }^{2}+{g}_{12}{%
\left\vert \psi _{+}\right\vert }^{2}\right) \psi _{-}+{\gamma }\psi _{-}%
\left[ \left\vert \psi _{-}(\mathbf{r})\right\vert ^{2}+\left\vert \psi _{+}(%
\mathbf{r})\right\vert ^{2}\right]   \notag \\
&&\times \ln \left[ \left\vert \psi _{-}(\mathbf{r})\right\vert
^{2}+\left\vert \psi _{+}(\mathbf{r})\right\vert ^{2}\right] +\kappa \psi
_{-}(\mathbf{r})\iint d\mathbf{r^{\prime }}R\left( \mathbf{r}-\mathbf{%
r^{\prime }}\right) \left[ \left\vert \psi _{-}\left( \mathbf{r^{\prime }}%
\right) \right\vert ^{2}+\left\vert \psi _{+}\left( \mathbf{r^{\prime }}%
\right) \right\vert ^{2}\right] ,  \label{Model2}
\end{eqnarray}%
where $\psi _{\pm }$ are wave functions of the two components, with density $%
\left\vert \psi _{+}(\mathbf{r})\right\vert ^{2}+\left\vert \psi _{-}(%
\mathbf{r})\right\vert ^{2}$ normalized to the above-mentioned reference
density $n_{0}$ (cf. Eq. (\ref{2D})), $\mathbf{r}$ = \{$x$,$y$\} is the set
of coordinates measured in units of $a_{\perp }$ (i.e., essentially,
measured in microns),  $\nabla ^{2}=\partial _{x}^{2}+\partial _{y}^{2}$,
the unit of time $t$ is $\sim 1$ ms, assuming that the mass of the particles
is $\sim 100$ proton masses, while $g_{11,22}>0$ and $g_{12}<0$, which are
proportional to the respective scattering lengths of the inter-particle
collisions, are strengths of the contact intra-component repulsion and
inter-component attraction, respectively. Further, coefficients ${\gamma }$
and $\kappa \sim Q^{2}/a_{\perp }^{3}$ (see Ref. \cite{QQI_LYY_PRA})
represent the LHY correction \cite{petrov2016} and MQQI, respectively. The
MQQI kernel is

\begin{equation}
R\left( \mathbf{r}-\mathbf{r^{\prime }}\right) =\frac{3-30\cos ^{2}\theta
+35\cos ^{4}\theta }{\left[ b^{2}+\left( \mathbf{r}-\mathbf{r^{\prime }}%
\right) ^{2}\right] ^{5/2}},  \label{kernel}
\end{equation}%
cf. Eq. (\ref{QQI_magn}), where $\cos ^{2}\theta =\left( x-x^{\prime
}\right) ^{2}/(\mathbf{r}-\mathbf{r^{\prime }})^{2}$ and $b\sim a_{\perp }$
is the cutoff, which is determined by the above-mentioned transverse
confinement size, that is set to be $1$, as said above.

Following Ref. \cite{petrov2016}, we focus the consideration on the
symmetric system, with $\psi _{+}=\psi _{-}\equiv \psi /\sqrt{2}$ and ${g}%
_{11}={g}_{22}=-{g}_{12}\equiv g$, thus reducing Eqs. (\ref{Model1}) and (%
\ref{Model2}) to a single equation in the finally scaled form:

\begin{equation}
i\frac{\partial }{\partial t}\psi =-\frac{1}{2}\triangledown ^{2}\psi +\ln
\left( \left\vert \psi \right\vert ^{2}\right) \cdot \left\vert \psi
\right\vert ^{2}\psi +\kappa \psi \int R\left( \mathbf{r}-\mathbf{r^{\prime }%
}\right) \left\vert \psi \left( \mathbf{r^{\prime }}\right) \right\vert ^{2}d%
\mathbf{r^{\prime }},  \label{model_single_re}
\end{equation}%
where $\kappa $ remains the single control parameter, while $b=1$ is fixed
in Eq. (\ref{kernel}), in accordance with what is said above. Dynamical
invariants of Eq. (\ref{model_single_re}) is the total norm,

\begin{equation}
N_{\text{2D}}=\int \left\vert \psi (\mathbf{r})\right\vert ^{2}d\mathbf{r,}
\label{norm}
\end{equation}%
which is proportional to the number of atoms in the BEC, and the total
energy, that includes the mean-field and LHY terms:%
\begin{equation}
E=\frac{1}{2}\int [\left\vert \triangledown \psi \right\vert ^{2}\mathbf{+}%
\left\vert \psi \right\vert ^{4}\ln \left( \frac{\left\vert \psi \right\vert
^{2}}{\sqrt{e}}\right) +\kappa \int R\left( \mathbf{r}-\mathbf{r^{\prime }}%
\right) ^{2}\left\vert \psi \left( \mathbf{r^{\prime }}\right) \right\vert
^{2}\left\vert \psi \left( \mathbf{r}\right) \right\vert ^{2}d\mathbf{%
r^{\prime }]}d\mathbf{r.}
\label{Energy}
\end{equation}%
Moving states also conserve the integral momentum, $\mathbf{P}=i\int \psi
\nabla \psi ^{\ast }d\mathbf{r}$, where $\ast $ stands for the complex
conjugate. Strictly speaking, the anisotropy present in Eqs. (\ref{kernel})
and (\ref{model_single_re}) breaks the conservation of the angular momentum
in the present setting. In fact, the QDs in our system are found to be
quasi-isotropic (as shown below), therefore the angular momentum is
approximately conserved.

The units of coordinates and time in Eq. (\ref{model_single_re}) are
tantamount, as said above, to $1$ $\mathrm{\mu }$m and $1$ ms, respectively,
while undoing the rescaling demonstrates that the value of the scaled norm $%
N=100$ corresponds to $\sim 5000$ particles in the underlying BEC.

We look for stationary solutions to Eq. (\ref{model_single_re}), which
represents stationary QDs with the chemical potential $\mu $ in the usual
form,
\begin{equation}
\psi \left( \mathbf{r},t\right) =\phi \left( \mathbf{r}\right) e^{-i\mu t},
\label{ansatz}
\end{equation}%
where $\phi \left( \mathbf{r}\right) $ is a wave function which is real for
fundamental states, and complex for vortical ones. Stationary QD solutions
were produced by means of the well-known imaginary-time-propagation (ITP)
method \cite{ITP1,ITP3,ITP2} applied to Eq. (\ref{model_single_re}).

\section{Stationary solutions for the quantum droplets (QDs)}

The strength of the MQQI, $\kappa $, and the total norm, $N$, are used as
control parameters in the present analysis. To generate stationary states,
the ITP method was initiated with the initial guess
\begin{equation}
\phi ^{\left( S\right) }\left( x,y\right) =Ar\exp \left( -\alpha
r^{2}-iS\theta \right) ,  \label{ansatz_S}
\end{equation}%
where $\alpha $ and $A$ are real constants, $r$ and $\theta $\ are 2D polar
coordinates, and integer $S$ is vorticity (topological charge, alias winding
number, $S=0$ corresponding to the ground state). Further, stability of the
stationary QD states was verified by direct simulations of the perturbed
real-time evolution. The simulations were run by dint of the split-step
Fourier-transform algorithm, adding random noise at the $1\%$ amplitude
level to the input, taking the input as $\psi =[1+\varepsilon ^{^{\prime
}}f(x,y)]\cdot \phi ^{\left( S\right) }\left( x,y\right) $, where $%
\varepsilon ^{^{\prime }}=1\%$ and $f(x,y)$ is a random function with the
value range $[0,1]$, which can be generated by the standard
\textquotedblleft rand" routine.

\subsection{Fundamental QDs}

\begin{figure}[h]
\includegraphics[width=0.5\columnwidth]{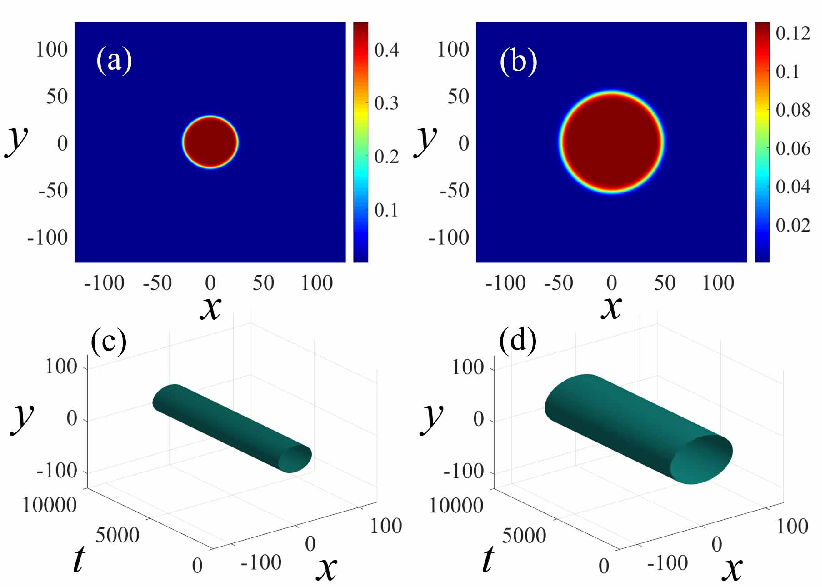}
% Here is how to import EPS art
\caption{Typical examples of stable quasi-isotropic fundamental quantum
droplets (QDs) produced by input~(\protect\ref{ansatz_S}) with $S=0$. Panels
(a) and (b) display the density profiles of the QDs with $(N,\protect\kappa %
)=(1000,0.1)$ and $(1000,0.5)$. (c) and (d): The perturbed evolution of the
QDs from (a) and (b), produced by simulations of Eq. (\protect\ref%
{model_single_re}) with the $1\%$ random noise added to the input.}
\label{fig:2}
\end{figure}
\ \

Typical examples of numerically constructed stable quasi-isotropic
fundamental QDs with $S=0$ are displayed in Fig. \ref{fig:2}. The parameters
are $(N,\kappa )=(1000,0.1)$ in (a) and $(1000,0.5)$ in (b). The stability
of these QDs was confirmed by direct simulations of their perturbed
evolution, as shown in Figs. \ref{fig:2}(c,d). Similar to QDs in binary
bosonic gases with contact-only interactions, the quasi-isotropic QDs in the
quadrupolar BEC generally feature flat-top density profiles. In this case,
one can apply the Thomas-Fermi (TF) approximation, neglecting the gradient
term in energy (\ref{Energy}). The corresponding approximation yields
\begin{equation}
E_{\mathrm{TF}}=\frac{1}{2}\left( \kappa \varepsilon n^{2}+n^{2}\ln \left(
\frac{n}{\sqrt{e}}\right) \right) A_{S},  \label{Eengry_TF}
\end{equation}%
where $n=\left\vert \psi \right\vert ^{2}$, $A_{S}=N/n$ is the effective
area of the flat-top QD, and
\begin{equation}
\varepsilon \equiv \int d\mathbf{r}R\left( \mathbf{r}\right) \approx 3.2,
\label{epsion}
\end{equation}%
with $R\left( \mathbf{r}\right) $ taken as per Eq. (\ref{kernel}), is an
effective measure of the repulsive nonlocality. Then, one can obtain the
equilibrium density $n_{e}$ and the effective area $A_{S}$ of the
quasi-isotropic QD from the energy-minimum condition $dE/dn=0$, which yields
\begin{eqnarray}
A_{S} &=&N\exp \left( \kappa \varepsilon +1/2\right) ,  \label{AS} \\
n_{e} &=&\exp \left( -\kappa \varepsilon -1/2\right) .  \label{ne}
\end{eqnarray}%
The chemical potential corresponding to the equilibrium density (\ref{ne}) is

\begin{equation}
\mu _{e}=\kappa \varepsilon n_{e}+n_{e}\ln n_{e}.  \label{mue}
\end{equation}

\begin{figure}[h]
\includegraphics[width=0.6\columnwidth]{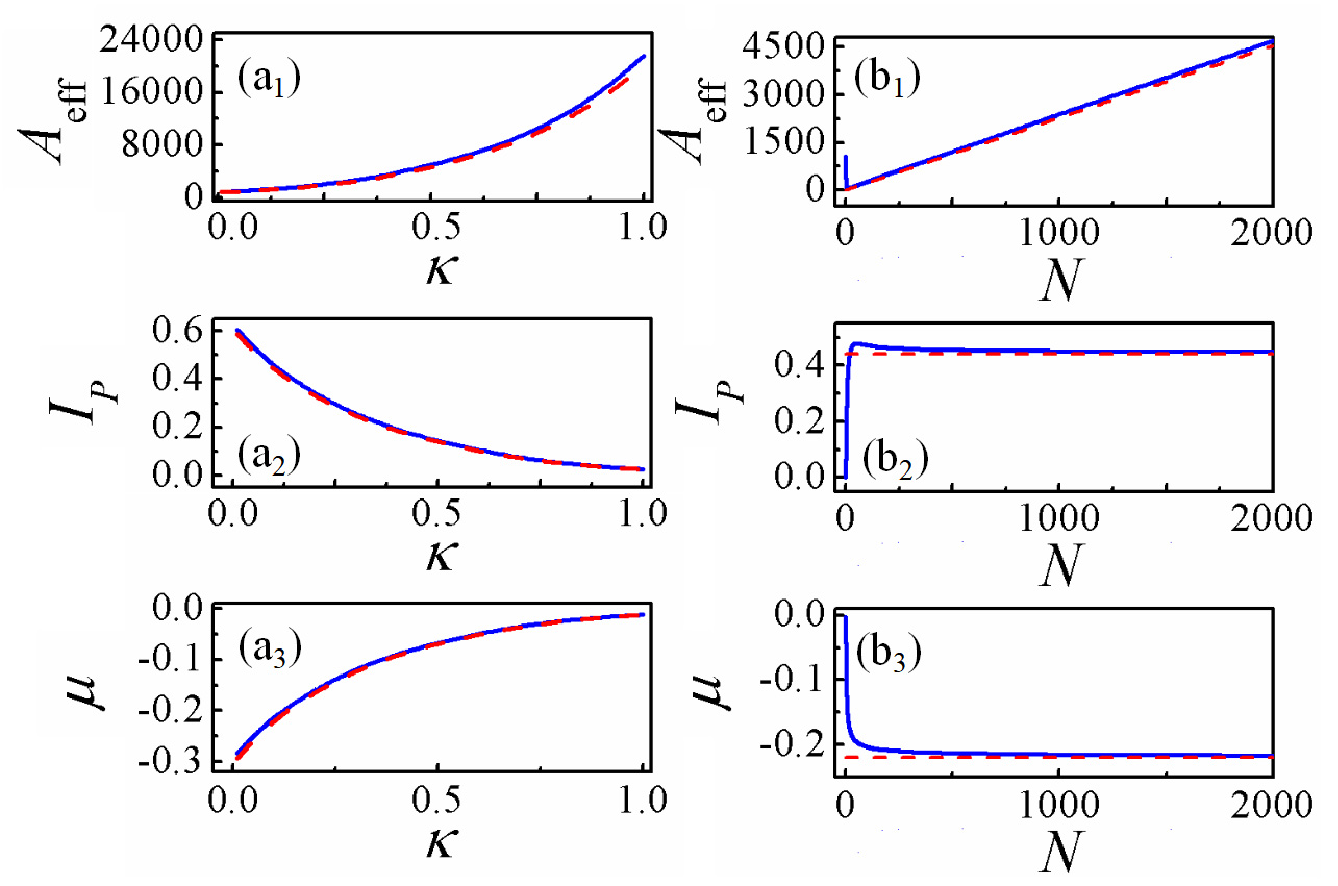}
\caption{The first column (a$_{1}$-a$_{3}$): the effective area ($A_{\text{%
eff}}$), peak density ($I_{p}$), and chemical potential ($\protect\mu $) of
the fundamental quasi-isotropic QDs as functions of the strength of $\protect%
\kappa $ with the norm $N=500$. The second column (b$_{1}$-b$_{3}$): the
dependence of $A_{\text{eff}}$, $I_{p}$ and $\protect\mu $ on the norm, $N$,
with $\protect\kappa =0.1$. Here, the solid blue line shows results of the
numerical simulations, and the dotted red line shows the analytical
approximation, see Eq. (\protect\ref{AS}), (\protect\ref{ne}), and (\protect
\ref{mue}), for $A_{\text{eff}}$, $I_{p}$ and $\protect\mu $, respectively. }
\label{fig3}
\end{figure}

Further, as a characteristics of the QDs, we define their effective area:
\begin{equation}
A_{\text{eff}}=\frac{\left( \int \left\vert \psi \right\vert ^{2}d\mathbf{r}%
\right) ^{2}}{\int \left\vert \psi \right\vert ^{4}d\mathbf{r}}.
\label{Aeff}
\end{equation}%
Comparing the results for $\kappa =0.1$ and $\kappa =0.5$ in Fig. \ref{fig:2}%
, we conclude that the effective area is larger in the latter case, as is
also seen in Fig. \ref{fig3}(a$_{1}$).

For the fixed norm, the dependence of the effective area $A_{\text{eff}}$ on
the MQQI strength $\kappa $ is shown in Fig. \ref{fig3}(a$_{1}$), which
indicates that the effective area exponentially grows the increase of $%
\kappa $. Here, the solid blue and dotted red lines show, respectively,
results of the numerical simulations and the analytical results, produced by
Eq. (\ref{AS}), the two curves being almost identical.

Next, we show the peak density, $I_{p}$ ($I_{p}=\left\vert \psi \right\vert
_{_{\max }}^{2}$), and the chemical potential $\mu $ of the stable
quasi-isotropic QDs as functions of the MQQI strength $\kappa $ in Figs. \ref%
{fig3}(a$_{2}$) and (a$_{3}$), respectively. Figure \ref{fig3}(a$_{2}$)
shows that $I_{p}$ decreases with $\kappa $, which is also in agreement with
Eq. (\ref{ne}), in which the peak density decays exponentially with the
increase of the MQQI strength.

In Figs. \ref{fig3}(b$_{1}$) - \ref{fig3}(b$_{3}$), we fix the MQQI strength
$\kappa $, to address the dependence of the effective area $A_{\text{eff}}$,
peak density $I_{p}$, and chemical potential $\mu $ on the total norm $N$.
Similar to the 2D anisotropic QDs in dipolar BECs, the effective area of the
quasi-isotropic QDs grows linearly with the increase of the total norm, see
Eq. (\ref{AS}), which is a natural feature of flat-top modes. In Fig. \ref%
{fig3}(b$_{2}$), the peak density saturates at $I_{p}\approx 0.448$, if $N$
is sufficiently large, which corroborates that the superfluid filling the
QDs is incompressible, as demonstrated first in Ref. \cite{petrov2015QDs}.
According to Eq. (\ref{ne}), the theoretical prediction for the equilibrium
density is about $0.440$, which is very close to the numerical results [see
the red dashed line in Fig. \ref{fig3}(b$_{2}$)]. The $\mu (N)$ curves in
Fig. \ref{fig3}(b$_{3}$) feature a negative slope, i.e., $d\mu /dN<0$,
satisfying the Vakhitov-Kolokolov (VK) criterion~\cite{VK}, which is a
necessary stability condition for self-trapped modes maintained by any
self-attractive nonlinearity. Figure \ref{fig3}(b$_{3}$) shows that the
chemical potential saturates at $\mu \approx -0.217$ when the norm is large.
The TF-predicted value given by Eq. (\ref{mue}) is $\mu _{e}\approx -0.220$
[see the red dashed line in Fig. \ref{fig3}(b$_{3}$)]. As shown in Fig. \ref%
{fig3}, all the numerical results agree well with the analytical predictions
provided by Eqs. (\ref{AS}), (\ref{ne}) and (\ref{mue}).

\subsection{QDs with embedded vorticity}

\begin{figure}[h]
\includegraphics[width=0.65\columnwidth]{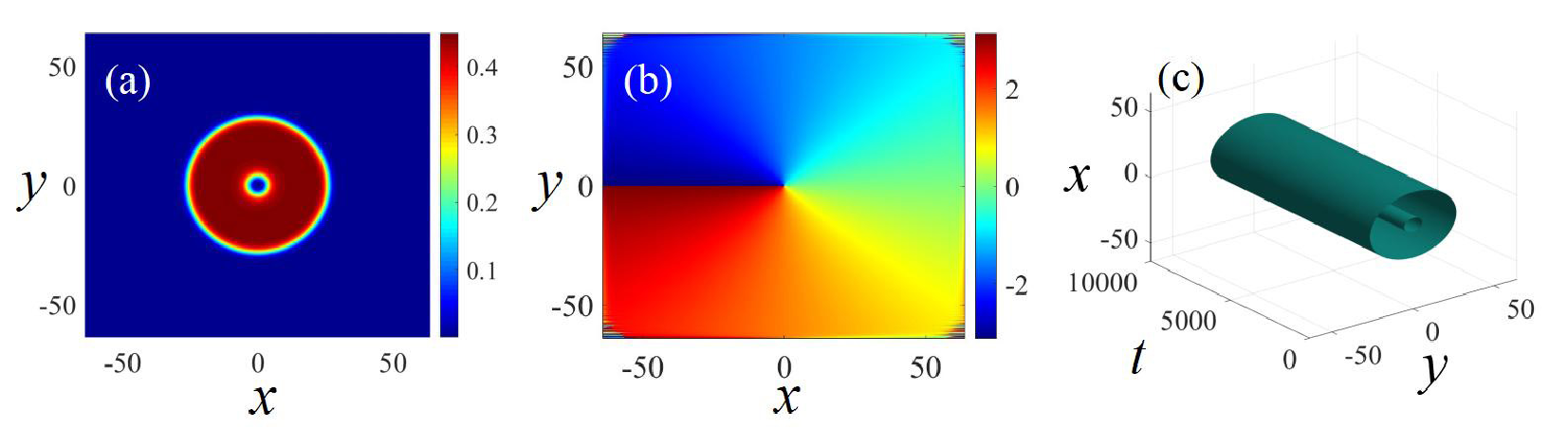}
% Here is how to import EPS art
\caption{(a) and (b): The density and phase patterns of a stable
quasi-isotropic VQD\ (vortex QD) produced by Eq.~(\protect\ref{ansatz_S})
with winding number $S=1$ and $(N,\protect\kappa )=(1000,0.1)$. (c)
Simulations of the perturbed evolution of the VQDs shown in panel (a) with $%
1\%$ random noise added to the input.}
\label{fig:4}
\end{figure}
\

The present system maintains stable vortex QDs (VQDs) as well. A typical
example of a numerically constructed one with winding number $S=1$, produced
by input~(\ref{ansatz_S}) with $S=1$, is displayed in Fig. \ref{fig:4}, for $%
N=1000$, and $\kappa =0.1$. The stability of these VQDs was confirmed by
direct simulations of its perturbed evolution, see Fig. \ref{fig:4}(c),
where $1\%$ random noise is added to the input.

To systematically study characteristics of the VQD families, we define their
aspect ratio and average angular momentum as follows:%
\begin{eqnarray}
A_{l} &=&\frac{W_{y}}{W_{x}},  \label{AL} \\
\bar{L}_{z} &=&\int \frac{\phi ^{\ast }\hat{L}_{z}\phi }{N}d\mathbf{r},
\label{LZ}
\end{eqnarray}%
where $W_{x,y}$ and $\bar{L}_{z}$ represent the effective lengths of the
VQDs in the $x/y$-direction, and the angular-momentum operator,
respectively:
\begin{eqnarray}
W_{x} &\equiv &\frac{\left( \int \left\vert \phi \left( x,y=0\right)
\right\vert ^{2}dx\right) ^{2}}{\int \left\vert \left( x,y=0\right)
\right\vert ^{4}dx},  \label{Wx} \\
W_{y} &\equiv &\frac{\left( \int \left\vert \phi \left( x=0,y\right)
\right\vert ^{2}dy\right) ^{2}}{\int \left\vert \left( x=0,y\right)
\right\vert ^{4}dy},  \label{Wy} \\
\hat{L}_{z} &=&-i\left( x\partial y-y\partial x\right) .  \label{Lz}
\end{eqnarray}%
In particular, $A_{l}<1$ indicates that the VQDs manifest anisotropy with
the elongation along the $x$-direction, while $A_{l}=1$ implies their
isotropy \cite{FOP_LGL}.

\begin{figure}[h]
\includegraphics[width=0.65\columnwidth]{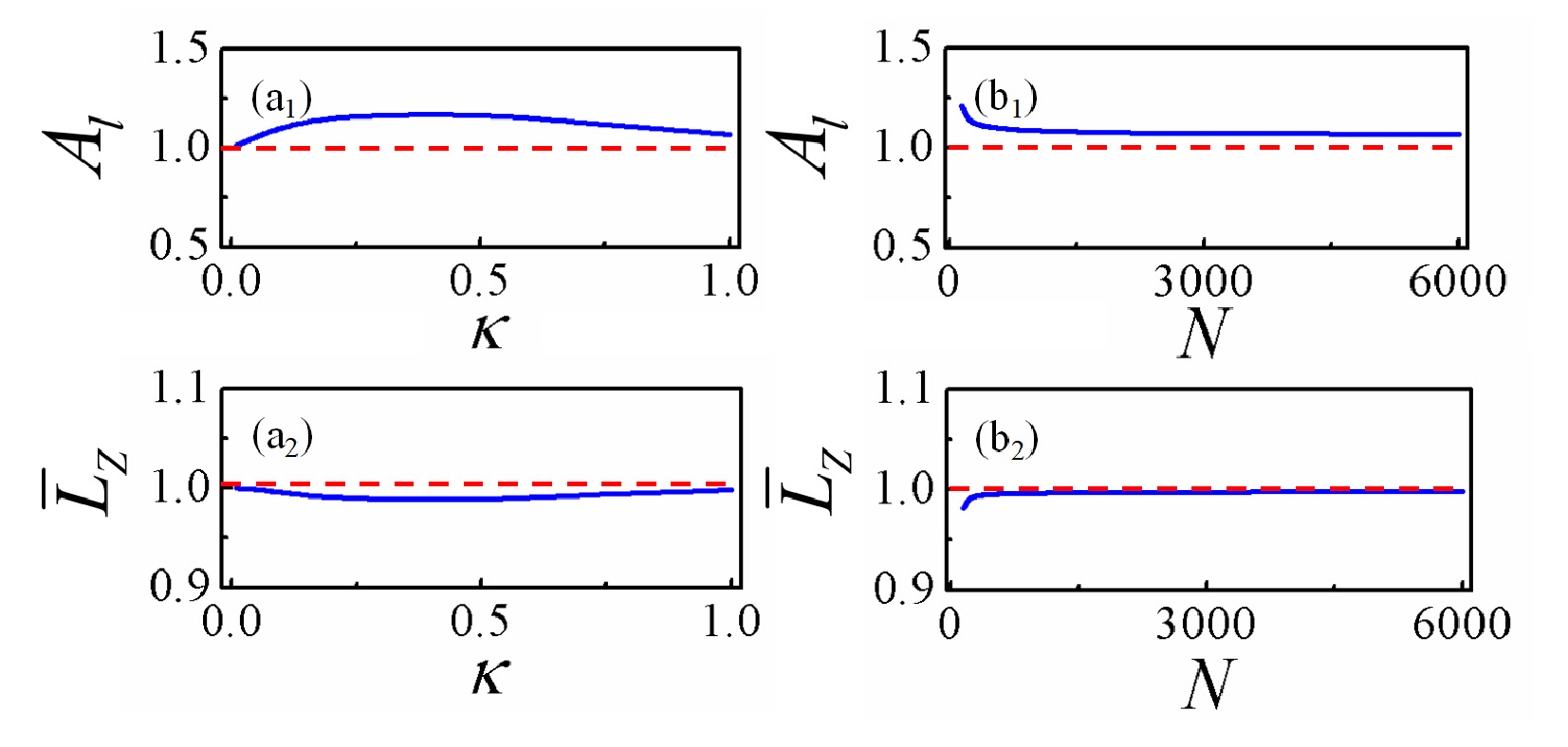}
% Here is how to import EPS art
\caption{Panels (a$_{1}$) and (a$_{2}$) display the aspect ratio $A_{l}$ and
average angular momentum $\bar{L}_{z}$ of stable quasi-isotropic VQDs with $%
N=500$, respectively, as functions of the MQQI strength $\protect\kappa $.
Panels (b$_{1}$) and (b$_{2}$) display the values, ($A_{l}$, $\bar{L}_{z}$),
for VQDs as functions of the norm $N$, with $\protect\kappa =0.1$. The red
dashed line designate the unit levels.}
\label{fig:5}
\end{figure}

The aspect ratio $A_{l}$ and average angular momentum $\bar{L}_{z}$ of
stable VQDs with $N=500$, defined as per Eqs. (\ref{AL})-(\ref{Lz}), are
displayed in Figs. \ref{fig:5}(a$_{1}$) and \ref{fig:5}(a$_{2}$),
respectively, as functions of the MQQI strength $\kappa $, varying in
interval $0<\kappa <1$ Note that, according to Fig. \ref{fig3}(a$_{1}$), the
effective area of QDs exponentially grows with the increase of $\kappa $,
making the calculations more difficult. For this reason, we here present the
results for $\kappa <1$. Figures \ref{fig:5}(b$_{1}$) and (b$_{2}$) display
the VQD characteristics $A_{l}$ and $\bar{L}_{z}$ as functions of the norm $N
$, with $\kappa =0.1$. In panels \ref{fig:5}(a$_{1}$), the aspect ratio
attains a maximum value $1.1742$ and approaches $1.0677$ at $\kappa
\rightarrow 1$. Figure \ref{fig:5}(a$_{2}$) depicts the respective orbital
momentum, whose average value is $0.9955$. In panels \ref{fig:5}(b$_{1}$)
and (b$_{2}$), the aspect ratio and orbital momentum have values of $%
A_{l}\approx 1.0921$ and $\bar{L}_{z}\approx 0.9957$, respectively. These
findings indicates that the system with the repulsive MQQI maintains stable
quasi-isotropic VQDs, whose aspect ratio is slightly larger than $1$. In
this connection, it is relevant to mention that for the fundamental QD modes
considered above, values of the aspect ratio are also approximately equal to
$1$, as the density distribution in those modes seem to be nearly
axisymmetric\ (isotropic).

\begin{figure}[h]
\includegraphics[width=0.65\columnwidth]{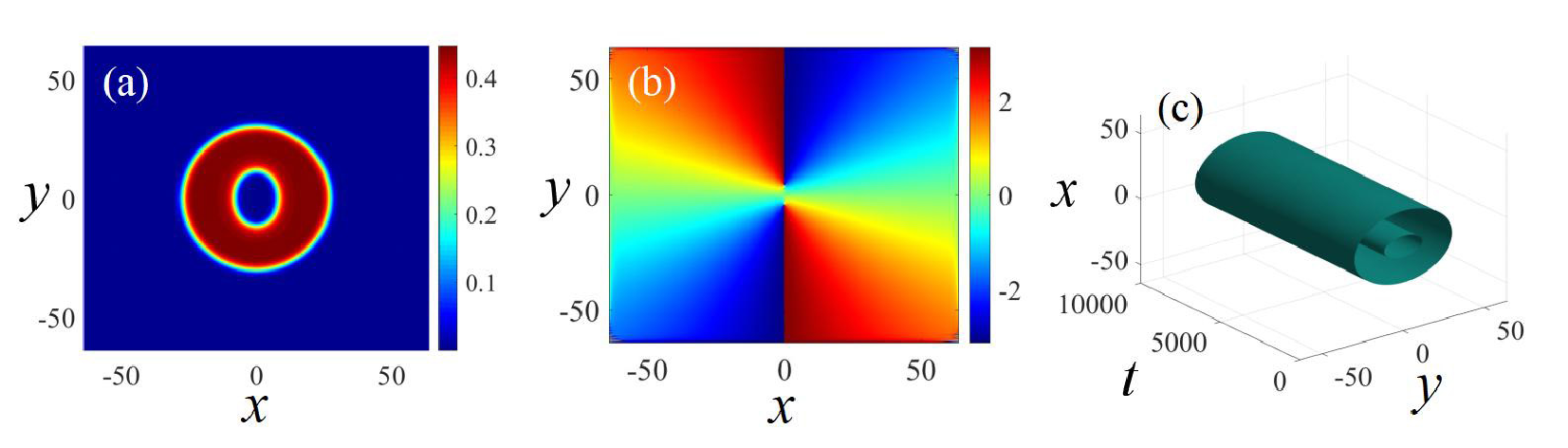}
% Here is how to import EPS art
\caption{Panels (a) and (b): The density and phase patterns of a stable VQD
with winding number $S=2$. The other parameters are\ the same as in Fig.
\protect\ref{fig:4}. (c) Simulations of the perturbed evolution of the VQD
shown in panels (a) and (b), with $1\%$ random noise added to the input.}
\label{fig:6}
\end{figure}

Simulations of in elastic collisions between moving stable VQDs with $S=1$
(see Fig. \ref{fig:9} below) demonstrate that they may merge into a single
breather with angular momentum (\ref{LZ}) close to $\bar{L}_{z}=2$. This
fact suggests that stable VQDs with $S=2$ may exist too. A typical example
of such a vortex state, produced by input Eq.~(\ref{ansatz_S}) with $S=2$,
is displayed in Fig. \ref{fig:6}, with the same parameters as in Fig. \ref%
{fig:4}. The respective values given by Eqs. (\ref{AL}) and (\ref{LZ}) for
this double VQD are $A_{l}=0.9718$ and $\bar{L}_{z}\approx 1.9824$ at $%
t=10000$.

\begin{figure}[h]
\includegraphics[width=0.65\columnwidth]{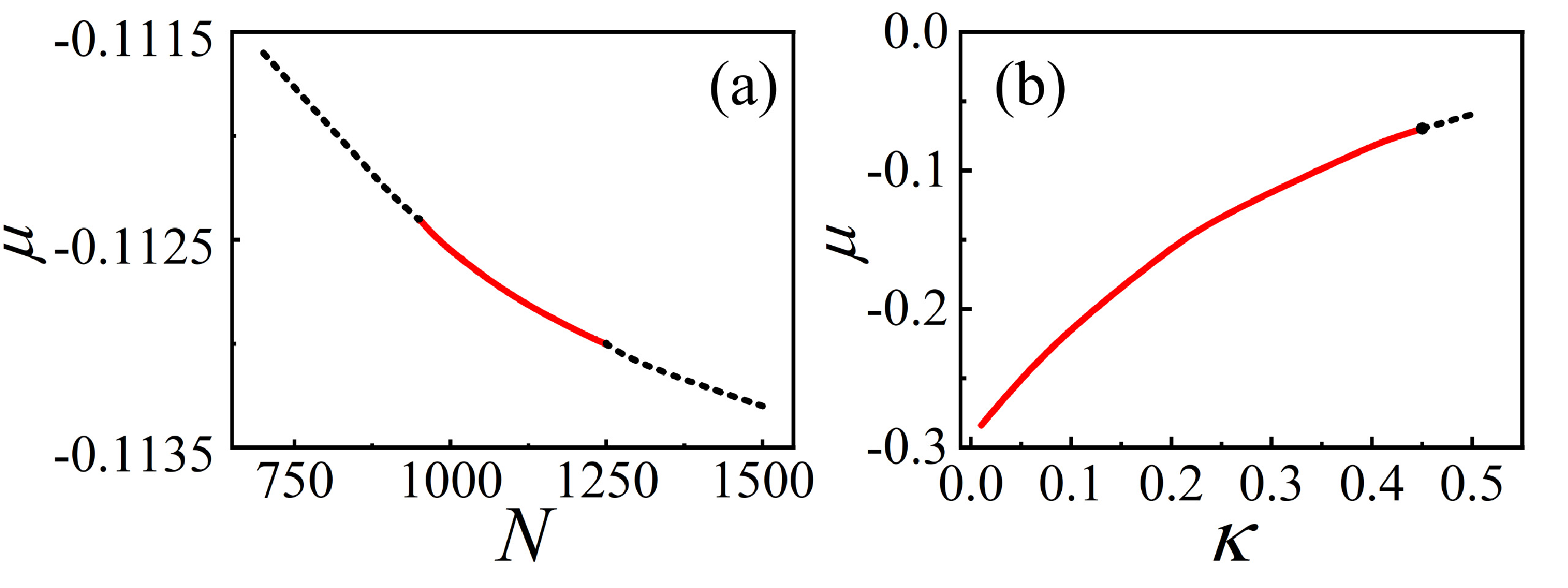}
% Here is how to import EPS art
\caption{(a) and (b): The chemical potential of the stable VQDs with $S=2$
versus $N$ (at $\protect\kappa =0.3$) and $\protect\kappa $ (at $N=1000$),
respectively. The red solid and black dotted segments of the curves
represent, respectively, stable and unstable states.}
\label{fig:7new}
\end{figure}

The family of VQDs with $S=2$ are characterized by dependences of their
chemical potential $\mu $ on $\kappa $ and $N$, as shown in Figs. \ref%
{fig:7new}(a) and (b).\ Next, we fix the characteristic value of the MQQI
strength, $\kappa =0.3$, to study a relation between the chemical potential $%
\mu $ of VQDs with $S=2$ and their norm $N$. Figure \ref{fig:7new}(a) shows
that the $\mu (N)$ curve has a negative slope, satisfying the VK criterion.
It is worthy to note that, for the current fixed values $S=2$ and $\kappa
=0.3$, the VQDs are stable in a finite interval, $950<N<1250$, see the red
line in Fig. \ref{fig:7new}(a), while the black dotted segments of the
curves represent unstable states, for the same parameters. In Fig. \ref%
{fig:7new}(b), with $N=1000$, the stable VQDs with $S=2$ populate the area
of $\kappa <0.45$.

\begin{figure}[h]
\includegraphics[width=0.9\columnwidth]{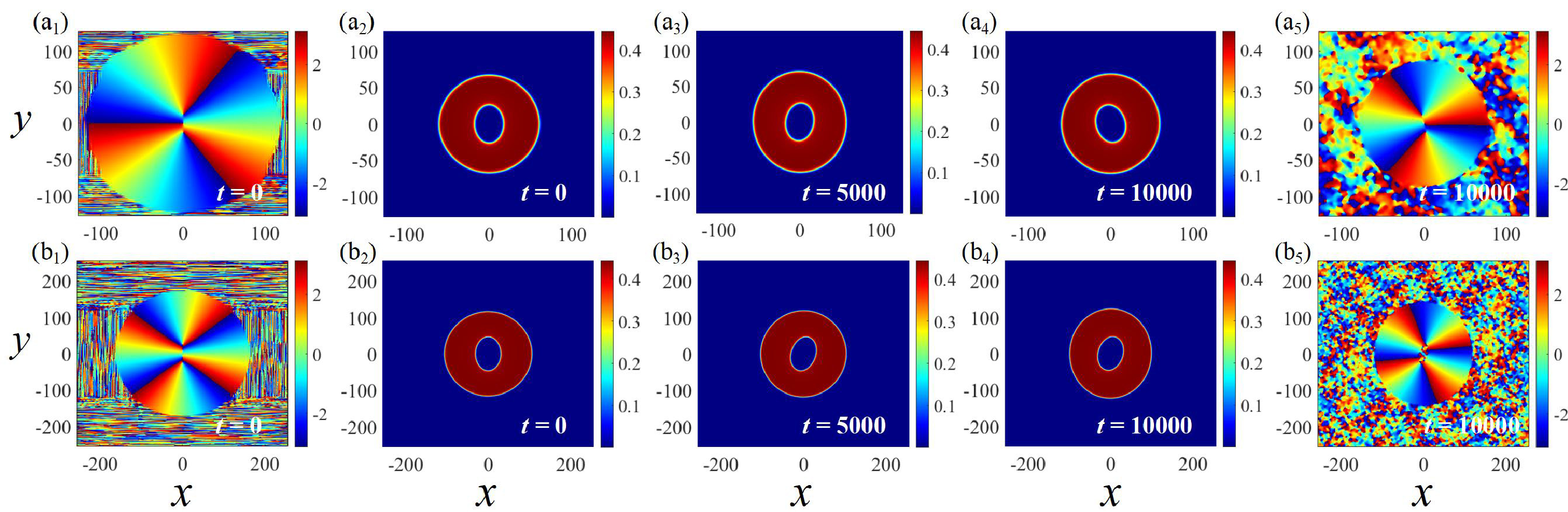}
% Here is how to import EPS art
\caption{Typical examples of quasi-isotropic VQDs with $(N,S)$ $=(5000,3)$
and $(N,S)=(15000,4)$. (a$_{1}$,b$_{1}$) and (a$_{5}$,b$_{5}$) display the
VQD phase patterns at $t=0$ and $t=10000$, respectively. Panels (a$_{2}$-a$%
_{4}$) and (b$_{2}$-b$_{4}$) show the density patterns of the VQDs at $t=0$,
$5000$, and $10000$, respectively.}
\label{fig:7}
\end{figure}

We have also constructed VQD states with higher vorticities, $S\geq 3$, as
shown in Fig. \ref{fig:7}. Typical examples of QDs with $S=3$ and $S=4$,
with parameters $(N,\kappa )=(5000,0.1)$ and $(N,\kappa )=(15000,0.1)$, are
displayed in Figs. \ref{fig:7}(a$_{1}$) - (a$_{5}$) and \ref{fig:7}(b$_{1}$)
- (b$_{5}$), respectively. Figures \ref{fig:7}(a$_{1}$,b$_{1}$) and (a$_{5}$%
,b$_{5}$), respectively, display the initial ($t=0$) and final ($t=10000$)
phase patterns of for $S=3$ and $4$. Figures \ref{fig:7}(a$_{2}$) - (a$_{4}$%
) and \ref{fig:7}(b$_{2}$) - (b$_{4}$) show the density patterns at $t=0$, $%
5000$, and $10000$. The simulations demonstrate that those QDs are stable
and feature quasi-isotropic shapes. According to Eq. (\ref{AL}), we have
found $A_{l}\approx 1.0308$ in Fig. \ref{fig:7}(a$_{4}$) and $A_{l}\approx
1.1338$ in Fig. \ref{fig:7}(b$_{4}$). Further, according to Eq. (\ref{LZ}),
we have calculated the average angular momentum of these QDs at $t=10000$, $%
\bar{L}_{z}\approx 2.9633$ in Fig. \ref{fig:7}(a$_{4}$) and $\bar{L}%
_{z}\approx 3.8869$ in Fig. \ref{fig:7}(b$_{4}$), which exhibit a small
deviation from the corresponding initial values of the angular momentum.

\section{Dynamics of the quantum droplets (QDs)}

It is well known that stable QDs in binary BECs can be set in motion by
opposite kicks with magnitude $\pm \eta $ applied along the $x$- or $y$%
-direction, which suggests to consider collisions between two QDs moving in
the opposite directions \cite{Grisha,PLA_HJZ}. The kicks can be readily applied
by optical pulses shone through the tapered solenoid.

Here, we address the collisions between the QDs moving in the $x$
direction. The corresponding initial states are constructed as
\begin{equation}
\psi \left( x,y,t=0\right) =\phi \left( x-x_{0},y\right) e^{-i\eta x}+\phi
\left( x+x_{0},y\right) e^{i\eta x},  \label{Dy_QIQD}
\end{equation}%
where $\phi \left( x\pm x_{0},y\right) $ represents the stationary
quasi-isotropic QDs initially centered at $x=\mp x_{0}$. It is necessary to
choose $x_{0}$ large enough so that the two QDs are initially placed far
from each other, avoiding any overlap.

\subsection{Collisions between fundamental QDs}

\begin{figure}[h]
\includegraphics[width=0.5\columnwidth]{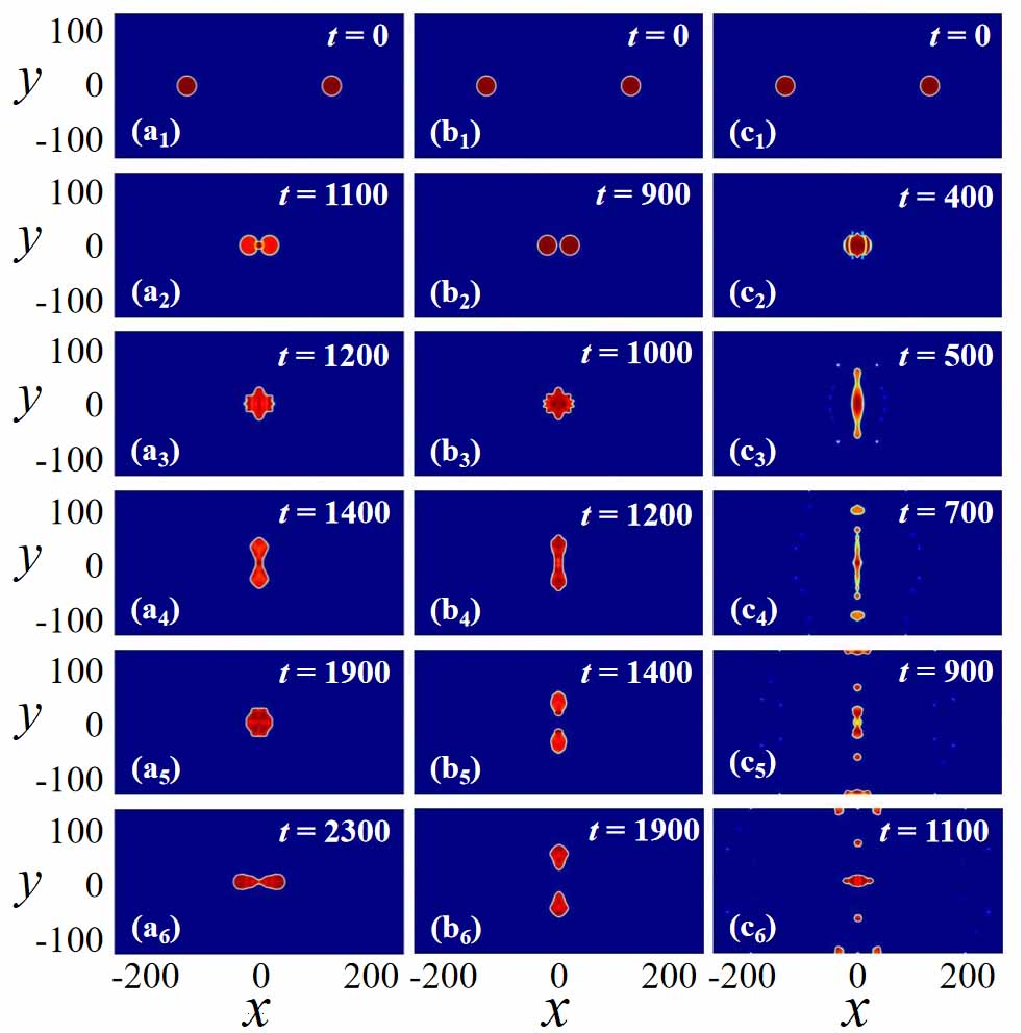}
% Here is how to import EPS art
\caption{Examples of collisions between moving quasi-isotropic fundamental
QDs, produced by input (\protect\ref{Dy_QIQD}) with $x_{0}=128$. Initial
parameters of the QDs are $\left( N,\protect\kappa \right) =\left(
500,0.05\right) $. (a$_{1}$) - (a$_{6}$): A strongly inelastic collision of
two QDs, set in motion by kicks $\pm 0.1$. (b$_{1}$) - (b$_{6}$): A
collision initiated by kicks $\pm 0.12$. (c$_{1}$) - (c$_{6}$): An inelastic
collision in the case of strong kicks, $\pm 0.3$, which implies the
collision with a large relative velocity. }
\label{fig:8}
\end{figure}

First, we consider the collisions between moving fundamental QDs ($S=0$). In
Ref. \cite{LB_PRA}, where collisions between two-component QDs have been
considered in the model with contact interactions, the simulations
demonstrated a trend to inelastic outcomes produced by the collisions
between QDs in the in-phase configuration. In Ref. \cite{Fop_zhangyb}, the
collisional dynamics has been studied for two symmetric QDs with equal
intra-species scattering lengths and equal densities of both components. In
Ref. \cite{FOP_LGL}, the collisions between moving 2D anisotropic vortex QDs
have been studied, demonstrating the formation of bound states with a
vortex-antivortex-vortex structure.

We have produced typical results for the collisions generated by input (\ref%
{Dy_QIQD}) with ($N,x_{0},\kappa $) $=$ ($500,128,0.05$). The results are
shown in Fig. \ref{fig:8}. Figures \ref{fig:8}(a$_{1}$) - (a$_{6}$) show the
result of the strongly inelastic collision between moving QDs set in motion
by kicks $\pm 0.1$. The result is fusion of the two fundamental QDs into a
quadrupole breather. It stretches periodically in the $x$ and $y$
directions, resembling the dynamics featured by a liquid drop \cite%
{Fop_zhangyb}. When the initial kicks increase to $\pm 0.12$, quasi-elastic
outcomes of the collision are observed. In this case, Figs. \ref{fig:8}(b$%
_{1}$) - (b$_{6}$) show that the colliding droplets separate into another
pair of droplets, slowly moving in the perpendicular direction, i.e., the
collision leads to the deflection of the direction of motion by 90$^{\text{o}%
}$.\ Similar results were mentioned in Refs.~\cite{photonics_YAW,Fop_zhangyb}%
. When the norm is fixed, the outcome of the collision is mainly determined
by the velocity, i.e., by kick $\eta $ \cite{LB_PRA,Fop_zhangyb,cnsns_LB}.
If the collision speed exceeds a critical value, an inelastic collision
takes place. A typical example of an inelastic collision of two
quasi-isotropic fundamental QDs is shown in Figs. \ref{fig:8}(c$_{1}$) - (c$%
_{6}$). After the collision, they split into several fragments. Note that
the norm and the MQQI strength (both of which affect the QD's size) also
affects the outcome of the collision. We do not consider these effects in
detail here.

\subsection{Collision between moving vortex quantum droplets (VQDs)}

\begin{figure}[h]
\includegraphics[width=0.5\columnwidth]{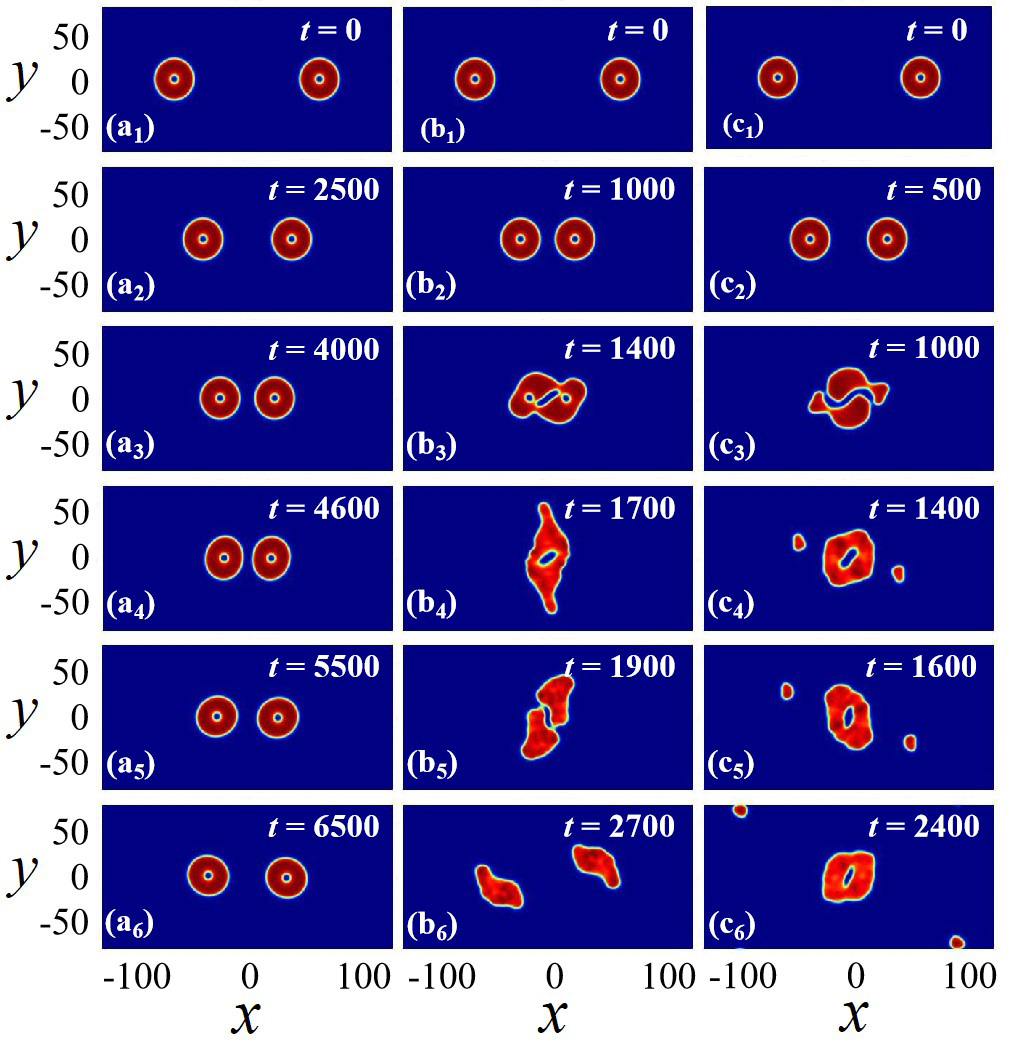}
% Here is how to import EPS art
\caption{Several typical examples of collisions between moving
quasi-isotropic vortex quantum droplets (VQDs), by inputting expression (%
\protect\ref{Dy_QIQD}) with $x_{0}=64$, which is selected with $\left( N,%
\protect\kappa ,S\right) =\left( 500,0.05,1\right) $. (a$_{1}$) - (a$_{6}$)
The contactless collision initiated by kick $\protect\eta =0.01$. (b$_{1}$)
- (b$_{6}$) Collision with the kick $\protect\eta =0.04$. (c$_{1}$) - (c$_{6}
$) Collision with the kick, $\protect\eta =0.06$. }
\label{fig:9}
\end{figure}

Next, we consider collisions between moving VQDs. For the input given by Eq.
(\ref{Dy_QIQD}) with ($N,x_{0},\kappa ,S$) $=$ ($500,64,0.05,1$), several
typical outcomes of the collision are presented in Fig. \ref{fig:9}. In
Figs. \ref{fig:9}(a$_{1}$- a$_{6}$), the collision with a small relative
speed, initiated by small kicks (here, they are $\pm 0.01$), demonstrates
that the VQDs bounce back, keeping their original vorticities, as a result
of the contactless collision. According to Eq. (\ref{LZ}), angular momenta
of the two droplets in Fig. \ref{fig:9}(a$_{6}$) are $\bar{L}_{z,\mathrm{%
right}}=$ $\bar{L}_{z,\mathrm{left}}\approx 1.005$. It is worth noting that
the mutual rebound does not occur in the in the case of the collision of
fundamental QDs, as seen in Fig. \ref{fig:8}. By gradually increasing the
kicks to $\pm 0.04$, the colliding VQDs merge and eventually split into a
couple of non-vortical fragments, see Fig. \ref{fig:9}(b$_{1}$) - (b$_{6}$).
The simulations demonstrate a trend to the merger of the colliding VQDs into
a vortex breather. In particular, this outcome, corresponding to the initial
kicks $\pm 0.06$, is observed in Figs. \ref{fig:9}(c$_{1}$) - \ref{fig:9}(c$%
_{6}$). The emerging breather absorbs $86.34\%$ of the total norm in Fig. %
\ref{fig:9}(c$_{6}$). The angular momentum of the breather is $\bar{L}%
_{z}\approx 1.6860$, which is relatively close to $2$. The VQDs being
quasi-isotropic, the average angular momentum is approximately conserved, as
mentioned above. This result suggests the existence of stable VQDs with
vorticity $S=2$. As demonstrated above, stable stationary VQDs withe $S=2$
are indeed available.

\section{Conclusion}

The purpose of this work is to establish the stability and characteristics
of 2D quasi-isotropic QDs (quantum droplets) in magnetic quadrupolar BECs.
The system is modeled by the coupled GP including the LHY (Lee-Huang-Yang)
terms, which represent the correction to the mean field theory produced by
quantum fluctuations. Stable quasi-isotropic QDs of the fundamental and
vortex types (with topological charges $S\leq 4$) have been produced by
means of imaginary-time simulations. The effect of the norm and strength of
the MQQI (magnetic quadrupole-quadrupole interactions), $\kappa $, on the
QDs were studied in detail. It was thus found that the QD's size increases
with the increase of $\kappa $. For stable QDs, the dependence between the
chemical potential and total norm obeys the VK (Vakhitov-Kolokolov)
criterion. The dependence of the effective size and peak density of the QDs
on $\kappa $ and norm $N$ have been studied too. The results of the
numerical simulation are consistent with the analytical predictions produced
by the TF (Thomas-Fermi) approximation. Typically, the predicted stable QDs
have the size in the range of $20-50$ $\mathrm{\mu }$m, with the number of
particles in the range of $10^{4}-10^{5}$, with density $\sim 10^{14}$ cm$%
^{-3}$.

Collisions between moving QDs have been also studied systematically, for the
fundamental and vortex droplets alike. For moving fundamental QDs, the
strongly inelastic, quasi-elastic, and inelastic outcomes of the collision
have been observed. For moving vortex QDs, a noteworthy results is the
contactless rebound of the colliding droplets.

The present analysis can be extended further. In particular, it is relevant
to look for essentially anisotropic QDs, especially higher-order vortical
ones, in the framework of the present model. A challenging option is to seek
for stable vortex QDs in the full three-dimensional setting including MQQI.

\section{Acknowledgments}

This work was supported by the Natural Science Foundation of GuangDong Province through grant Nos. 2024A1515030131, 2021A1515010214, GuangDong Basic and Applied Basic Research Foundation through grant No. 2023A1515110198, National Natural Science Foundation of China through grants Nos. 12274077 and 11905032, the Research Fund of the Guangdong-Hong Kong-Macao Joint Laboratory for Intelligent Micro-Nano Optoelectronic Technology through grant No. 2020B1212030010, and Israel Science Foundation through Grant No. 1695/22.


\begin{thebibliography}{999}
\bibitem{petrov2015QDs} Petrov D S 2015 \emph{Phys. Rev. Lett.} \textbf{115}
155302

\bibitem{LHY} Lee T D, Huang K S and Yang C N 1957 \emph{Phys. Rev.} \textbf{%
106} 1135

\bibitem{QD_K} Cabrera C R, Tanzi L, Sanz J, Naylor B, Thomas P, Cheiney P
and Tarruell L 2017 \emph{Science} \textbf{359} 301

\bibitem{QD_Tarruell_2} Cheiney P, Cabrera C R, Sanz J, Naylor B, Tanzi L
and Tarruell L 2018 \emph{Phys. Rev. Lett.} \textbf{120} 135301

\bibitem{PRL120_235301} Semeghini G, Ferioli G, Masi L, Mazzinghi C,
Wolswijk L, Minardi F, Modugno M, Modugno G, Inguscio M and Fattori M 2018
\emph{Phys. Rev. Lett.} \textbf{120} 235301

\bibitem{QD_KNA_WDJ} Guo Z, Jia F, Li L, Ma Y, Hutson J M, Cui X and Wang D
2021 \emph{Phys. Rev. Res.} \textbf{3} 033247

\bibitem{QD_dy_nature} Schmitt M, Wenzel M, B\"{o}ttcher F, Ferrier-Barbut I
and Pfau T 2016 \emph{Nature (London)} \textbf{539} 259

\bibitem{QD_er_prx} Chomaz L, Baier S, Petter D, Mark M J, W\"{a}chtler F,
Santos L and Ferlaino F 2016 \emph{Phys. Rev. X} \textbf{6} 041039

\bibitem{petrov2016} Petrov D S and Astrakharchik G E 2016 \emph{Phys. Rev.
Lett.} \textbf{117} 100401

\bibitem{PRA103_033312} Lavoine L and Bourdel T 2021 \emph{Phys. Rev. A}
\textbf{103} 033312

\bibitem{PRL120_160402} Ferrier-Barbut I, Wenzel M, B{\"{o}}ttcher F, Langen
T, Isoard M, Stringari S and Pfau T 2018 \emph{Phys. Rev. Lett.} \textbf{120}
160402

\bibitem{PRA98_033612} Boudjem\^{a}a A 2018 \emph{Phys. Rev. A} \textbf{98}
033612

\bibitem{PRA_HH} Hu H and Liu X 2020 \emph{Phys. Rev. A} \textbf{102} 043302

\bibitem{PRR4_013168} Pylak M, Gampel F, P\l odzie\'{n} M and Gajda M 2022
\emph{Phys. Rev. Res.} \textbf{4} 013168

\bibitem{PRA102_023318} Parisi L and Giorgini S 2020 \emph{Phys. Rev. A}
\textbf{102} 023318

\bibitem{dong2022bistable} Dong L, Liu D, Du Z, Shi K and Qin W 2022 \emph{%
Phys. Rev. A} \textbf{105} 033321

\bibitem{PRA103_013312} Zin P, Pylak M and Gajda M 2021 \emph{Phys. Rev. A}
\textbf{103} 013312

\bibitem{PRE102_062217} Otajonov S R, Tsoy E N and Abdullaev F Kh 2020 \emph{%
Phys. Rev. E} \textbf{102} 062217

\bibitem{PRA101_051601(R)} Tylutki M, Astrakharchik G E, Malomed B A and
Petrov D S 2020 \emph{Phys. Rev. A} \textbf{101} 051601(R)

\bibitem{FOP_LZH} Luo Z, Pang W, Liu B, Li Y and Malomed B A 2021 \emph{%
Front. Phys.} \textbf{16} 32201

\bibitem{CXL_PRA2018} Cui X 2018 \emph{Phys. Rev. A} \textbf{98} 023630

\bibitem{PRResearch2_043074} Wang Y, Guo L, Yi S and Shi T 2020 \emph{Phys.
Rev. Res.} \textbf{2} 043074

\bibitem{PRL126_244101} Dong L and Kartashov Y V 2021 \emph{Phys. Rev. Lett.}
\textbf{126} 244101

\bibitem{CSF_WHC} Huang H, Wang H, Chen M, Lim C S and Wong K 2022 \emph{%
Chaos, Solitons and Fractals} \textbf{158} 112079

\bibitem{FOP_GMY} Guo M and Pfau T 2021 \emph{Front. Phys.} \textbf{16} 32202

\bibitem{FOP_Boris} Malomed B A 2021 \emph{Front. Phys.} \textbf{16} 22504

\bibitem{Nonlinear Dyn. Zhou 2022} Zhou Z, Shi Y, Ye F, Chen H, Tang S, Deng
H and Zhong H 2022 \emph{Nonlinear Dyn.} \textbf{110} 3769

\bibitem{CSF Zhou 2021} Zhou Z, Shi Y, Tang S, Deng H, Wang H, He X and
Zhong H 2021 \emph{Chaos, Solitons and Fractals} \textbf{150} 111193

\bibitem{NJP1} Kartashov Y V, Lashkin V M, Modugno M and Torner L 2022 \emph{%
New J. Phys.} \textbf{24} 073012

\bibitem{NJP5} Zin P, Pylak M and Gajda M 2021 \emph{New J. Phys.} \textbf{23%
} 033022

\bibitem{NJP6} Cikojevi\'{c} V, Marki\'{c} L V and Boronat J 2020 \emph{New
J. Phys.} \textbf{22} 053045

\bibitem{NJP7} Cidrim A, Salasnich L and Macr\'{\i} T 2021 \emph{New J. Phys.%
} \textbf{23} 023022

\bibitem{NJP8} Karpiuk T, Gajda M and Brewczyk M 2020 \emph{New J. Phys.}
\textbf{22} 103025

\bibitem{cpb5} Zhang F and Yin L 2022 \emph{Chin. Phys. Lett.} \textbf{39}
060301

\bibitem{cpb_yw} Wang J, Pan J, Cui X and Yi W 2020 \emph{Chin. Phys. Lett.}
\textbf{37} 076701

\bibitem{cpb_lxj} Wang J, Liu X and Hu H 2021 \emph{Chin. Phys. B} \textbf{30%
} 010306

\bibitem{PRl_122_090401} Ferioli G, Semeghini G, Masi L, Giusti G, Modugno
G, Inguscio M, Gallem\'{\i} A, Recati A and Fattori M 2019 \emph{Phys. Rev.
Lett.} \textbf{122} 090401

\bibitem{CSF_ZFY} Zhao F, Yan Z, Cai X, Li C, Chen G, He H, Liu B and Li Y
2021 \emph{Chaos, Solitons and Fractals} \textbf{152} 111313

\bibitem{RPP_dipolar} B\"{a}ttcher F, Schmidt J N, Hertkorn J, Ng K S H,
Graham S D, Guo M, Langen T and Pfau T 2021 \emph{Rep. Prog. Phys.} \textbf{%
84} 012403

\bibitem{PRS} D'Errico C, Burchianti A, Prevedelli M, Salasnich L, Ancilotto
F, Modugno M, Minardi F and Fort C 2019 \emph{Phys. Rev. Res.} \textbf{1}
033155

\bibitem{JPBmagnetic} Ferrier-Barbut I, Schmitt M, Wenzel M, Kadau H and
Pfau T 2016 \emph{J. Phys. B: At. Mol. Opt. Phys.} \textbf{49} 214004

\bibitem{Fop_zhangyb} Hu Y, Fei Y, Chen X and Zhang Y 2022 \emph{Front. Phys.%
} \textbf{17} 61505

\bibitem{zhangYC2018long} Zhang Y, Walther V and Pohl T 2018 \emph{Phys.
Rev. Lett.} \textbf{121} 073604

\bibitem{zhangYC2023quantum} Dong B, Zhang Y and Zhang X 2023 \emph{Optik}
\textbf{273} 170484

\bibitem{zhangYC2021self} Zhang Y, Walther V and Pohl T 2021 \emph{Phys.
Rev. A} \textbf{103} 023308

\bibitem{Yin2021} Xiong Y and Yin L 2021 \emph{Chin. Phys. Lett.} \textbf{38}
070301

\bibitem{chen2021one} Chen J and Zeng J 2021 \emph{Results Phys.} \textbf{21}
103781

\bibitem{ma2021borromean} Ma Y, Peng C and Cui X 2021 \emph{Phys. Rev. Lett.}
\textbf{127} 043002

\bibitem{cui2021droplet} Cui X and Ma Y 2021 \emph{Phys. Rev. Res.} \textbf{3%
} L012027

\bibitem{xu2022three} Xu S, Lei Y, Du J, Zhao Y, Hua R and Zeng J 2022 \emph{%
Chaos Solitons Fractals} \textbf{164} 112665

\bibitem{zhou2019dynamics} Zhou Z, Yu X, Zhou Y and Zhong H 2019 \emph{%
Commun Nonlinear Sci Numer Simul} \textbf{78} 104881

\bibitem{wang2020thermal} Wang J, Hu H and Liu X 2020 \emph{New J. Phys.}
\textbf{22} 103044

\bibitem{boudjemaa2023quantum} Boudjem{\^a}a A and Abbas K 2023 \emph{New J.
Phys.} \textbf{25} 093052

\bibitem{rakshit2019self} Rakshit D, Karpiuk T, Zin P, Brewczyk M,
Lewenstein M and Gajda M 2019 \emph{New J. Phys.} \textbf{21} 073027

\bibitem{pshenichnyuk2017static} Pshenichnyuk I A 2017 \emph{New J. Phys.}
\text{19} 105007

\bibitem{zin2022self} Zin P, Pylak M, Idziaszek Z and Gajda M 2022 \emph{New
J. Phys.} \textbf{24} 113038

\bibitem{Fischer2006} Sch\"{u}tzhold R, Uhlmann M, Xu Y and Fischer U R 2006
\emph{Int. J. Mod. Phys. B} \textbf{20} 3555

\bibitem{LYY_SOC_QD} Li Y, Luo Z, Liu Y, Chen Z, Huang C, Fu S, Tan H and
Malomed B A 2017 \emph{New J. Phys.} \textbf{19} 113043

\bibitem{zhangYC2021phases} Zhang Y, Pohl T and Maucher F 2021 \emph{Phys.
Rev. A} \textbf{104} 013310

\bibitem{prx_Supersolid} B\"{o}ttcher F, Schmidt J, Wenzel M, Hertkorn J,
Guo M, Langen T and Pfau T 2019 \emph{Phys. Rev. X} \textbf{9} 011051

\bibitem{metastability} Kartashov Y V, Malomed B A and Torner L 2019 \emph{%
Phys. Rev. Lett.} \textbf{122} 193902

\bibitem{PRL_LYY2019} Zhang X, Xu X, Zheng Y, Chen Z, Liu B, Huang C,
Malomed B A and Li Y 2019 \emph{Phys. Rev. Lett.} \textbf{123} 133901

\bibitem{LB_PRA} Liu B, Zhang H, Zhong R, Zhang X, Qin X, Huang C, Li Y and
Malomed B A 2019 \emph{Phys. Rev. A} \textbf{99} 053602

\bibitem{Fetter} Fetter A 2009 \emph{Rev. Mod. Phys}. \textbf{81} 647

\bibitem{BEC_OL} Morsch O and Oberthaler M 2006 \emph{Rev. Mod. Phys.}
\textbf{78} 179

\bibitem{cpb3} Liu Y and Yang S 2014 \emph{Chin. Phys. B} \textbf{23} 110308

\bibitem{cpb1} Yang R and Yang J 2008 \emph{Chin. Phys. B} \textbf{17} 1189

\bibitem{cpb_wlh} Wang Q, Yang H, Su N and Wen L 2020 \emph{Chin. Phys. B}
\textbf{29} 116701

\bibitem{cpl_dcq} Cao Q and Dai C 2021 \emph{Chin. Phys. Lett.} \textbf{38}
090501

\bibitem{cpl_xsl} Li B, Zhao Y, Xu S, Zhou Q, Fu Q, Ye F, Hua C, Chen M, Hu
H, Zhou Q and Qiu Z 2023 \emph{Chin. Phys. Lett.} \textbf{40} 044201

\bibitem{RFZ2} Zhang R, Zhang X and Li L 2019 \emph{Phys. Lett. A} \textbf{%
383} 231

\bibitem{WJG} Wang J, Wang W, Bai X and Yang S 2019 \emph{Eur. Phys. J. Plus}
\textbf{134} 27

\bibitem{ZXFPRA95} Zhang X, Kato M, Han W, Zhang S and Saito H 2017 \emph{%
Phys. Rev. A} \textbf{95} 033620

\bibitem{Brtka} Brtka M, Gammal A and Malomed B A 2010 \emph{Phys. Rev. A}
\textbf{82} 053610

\bibitem{HV_WB} Wen L, Xiong H and Wu B 2010 \emph{Phys. Rev. A} \textbf{82}
053627

\bibitem{HV_pla} Subramaniyan S 2017 \emph{Phys. Lett. A} \textbf{381} 3062

\bibitem{PRA_HCQ2017} Huang C, Lyu L, Huang H, Chen Z, Fu S, Tan H, Malomed
B A and Li Y 2017 \emph{Phys. Rev. A} \textbf{96} 053617

\bibitem{FOP_ZYY} Zheng Y, Chen S, Huang Z, Dai S, Liu B, Li Y and Wang S
2021 \emph{Front. Phys.} \textbf{16} 22501

\bibitem{NJP_LB2022} Liu B, Chen Y, Yang A, Cai X, Liu Y, Luo Z, Qin X,
Jiang X, Li Y and Malomed B A 2022 \emph{New J. Phys.} \textbf{24} 123026

\bibitem{PRE_LB} Liu B, Cai X, Qin X, Jiang X, Xie J, Malomed B A and Li Y
2023 \emph{Phys. Rev. E} \textbf{108} 044210

\bibitem{QJL_PRA} Qin J, Dong G and Malomed B A 2016 \emph{Phys. Rev. A}
\textbf{94} 053611

\bibitem{Ben Li} Sakaguchi H Li B and Malomed B A 2014 \emph{Phys. Rev. E}
\textbf{89}, 032920

\bibitem{Han Pu} Zhang Y-C, Zhou Z-W, Malomed B\ A and Pu H 2015 \emph{Phys.
Rev. Lett.} \textbf{115} 253902

\bibitem{2Ddipolar_njp} Boudjem\^{a}a A 2019 \emph{New J. Phys.} \textbf{21}
093027

\bibitem{pragroundstate} W\"{a}chtler F and Santos L 2016 \emph{Phys. Rev. A}
\textbf{94} 043618

\bibitem{dip_PRA-unstable} Cidrim A, dos Santos F E A, Henn E A L and Macr%
\'{\i} T 2018 \emph{Phys. Rev. A} \textbf{98} 023618

\bibitem{photonics_YAW} Yang A, Li G, Jiang X, Fan Z, Chen Z, Liu B and Li Y
2023 \emph{Photonics} \textbf{10} 405

\bibitem{FOP_LGL} Li G, Jiang X, Liu B, Chen Z, Malomed B A and Li Y 2024
\emph{Front. Phys.} \textbf{19} 22202

\bibitem{FOP_QQI_LYY} Huang J, Jiang X, Chen H, Fan Z, Pang W and Li Y 2015
\emph{Front. Phys.} \textbf{10} 1

\bibitem{QQI_LYY_PRA} Li Y, Liu J, Pang W and Malomed B A 2013 \emph{Phys.
Rev. A} \textbf{88} 063635

\bibitem{ZRX_QQI} Zhong R, Huang N, Li H, He H, L\"{u} J, Huang C and Chen Z
2018 \emph{Int. J. Mod. Phys. B} \textbf{32} 1850107

\bibitem{mishra2020self} Mishra C, Santos L and Nath R 2020 \emph{Phys. Rev.
Lett.} \textbf{124} 073402

\bibitem{ghosh2022droplet} Ghosh R, Mishra C, Santos L and Nath R 2022 \emph{%
Phys. Rev. A} \textbf{106} 063318

\bibitem{molecules} Junquera-Hern\'{a}ndez J M, S\'{a}nchez-Mar\'{\i}n J and
Maynau D 2002 \emph{Chem. Phys. Lett.} \textbf{359} 343-348

\bibitem{diatoms} Bhongale S G, Mathey L, Zhao E, Yelin S F and Lemeshko M
2013 \emph{Phys. Rev. Lett.} \textbf{110} 155301

\bibitem{CGH_QQI_CNCNS} Chen G, Liu Y and Wang H 2017 \emph{Commun.
Nonlinear Sci. Numer. Simulat} \textbf{48} 318

\bibitem{Pit} Pitaevskii L P and Stringari S 2003 \textit{Bose-Einstein
Condensation} (Oxford University Press, Oxford)

\bibitem{cpb_yisu_QQI} Wang A and Yi S 2018 \emph{Chin. Phys. B} \textbf{27}
120307

\bibitem{smith2021quantum} Smith J C, Baillie D and Blakie P B 2021 \emph{%
Phys. Rev. Lett.} \textbf{126} 025302

\bibitem{bisset2021quantum} Bisset R N, Pe\~{n}a Ardila L A and Santos L
2021 \emph{Phys. Rev. Lett.} \textbf{126} 025301

\bibitem{ITP1} Chiofalo M L, Succi S and Tosi M P 2000 \emph{Phys. Rev. E}
\textbf{62} 7438

\bibitem{ITP3} Bao W and Du Q 2004 \emph{SIAM J. Sci. Comp.} \textbf{25}
1674-1697

\bibitem{ITP2} Yang J and Lakoba T I 2008 \emph{Stud. Appl. Math.} \textbf{%
120} 265

\bibitem{VK} Vakhitov N G and Kolokolov A A 1973 \emph{Radiophys Quantum
Electron.} \textbf{16} 783

\bibitem{Grisha} Astrakharchik G E and Malomed B A 2018 \emph{Phys. Rev. A}
\textbf{98} 013631

\bibitem{PLA_HJZ} Hu J, Li S, Chen Z, L\"{u} J, Liu B and Li Y 2020 \emph{%
Phys. Lett. A} \textbf{384} 126448

\bibitem{cnsns_LB} Lin Z, Xu X, Chen Z, Yan Z, Mai Z and Liu B 2021 \emph{%
Commun. Nonlinear Sci. Numer. Simulat.} \textbf{93} 105536
\end{thebibliography}
\end{document}